\documentclass[fleqn,usenatbib]{mnras}

\usepackage{newtxtext,newtxmath}

\usepackage[T1]{fontenc}
\usepackage{ae,aecompl}
\usepackage{xcolor}

\usepackage{graphicx}	
\usepackage{amsmath}	
\usepackage{amssymb}	

\title[Statistical analysis of emerging ARs]{Extended statistical analysis of emerging solar active regions}

\author[A.S. Kutsenko, V.I. Abramenko and A.A. Pevtsov]{
Alexander S. Kutsenko,$^{1}$\thanks{E-mail: alex.s.kutsenko@gmail.com (ASK)}
Valentina I. Abramenko$^{1}$
and Alexei A. Pevtsov$^{2}$
\\
$^{1}$Crimean Astrophysical Observatory, p/o Nauchny, Crimea, 298409, Russia\\
$^{2}$National Solar Observatory, Boulder, CO, 80303, USA\\
}

\date{Accepted XXX. Received YYY; in original form ZZZ}

\pubyear{2018}

\begin{document}
\label{firstpage}
\pagerange{\pageref{firstpage}--\pageref{lastpage}}
\maketitle

\begin{abstract}
We use observations of line-of-sight magnetograms from Helioseismic and 
Magnetic Imager (HMI) on board of Solar Dynamics Observatory (SDO) to 
investigate polarity separation, magnetic flux, flux emergence rate, twist and tilt 
of solar emerging active regions. 
Functional dependence of polarity 
separation and maximum magnetic flux of 
an active region is in agreement with 
a simple model of flux emergence as the 
result of buoyancy forces. Our investigation did not reveal any strong 
dependence of emergence rate on twist properties of active regions.
\end{abstract}




\section{Introduction}

It is widely accepted now that solar active regions (ARs) are a manifestation of emerged portions of subphotospheric toroidal magnetic flux \citep{Parker1955}. The magnetic field that forms active regions is generated somewhere in the convection zone (CZ) in the processes collectively called the solar dynamo. The conditions in the tachocline at the base of the CZ allow the magnetic flux being amplified by stretching and wrapping of magnetic field lines by the differential rotation (the so-called ``$\Omega$ effect'') and the action of turbulent convection ($\alpha$-effect) \citep[e.g.,][]{Babcock1961,Ossendrijver2003,Charbonneau2005,Charbonneau2010,Charbonneau2014,Brun2015,Ferriz-Mas1994,Ferriz-Mas2007}.
In addition to amplification, the magnetic field get organized into a concentrations referred to as flux tubes, in which the individual field lines are linked together by wrapping around other field lines. While the exact mechanisms leading to formation of flux tubes are not well-understood yet, the concept is used widely in representing the 
dynamics and properties of strong magnetic fields on the Sun. After the flux tube is formed, one of its portion 
may become unstable and start rising to the surface due to buoyancy \citep{Parker1975,Fan2004}. The process starts when the magnetic pressure inside the magnetic flux tube becomes comparable to the gas pressure, i.e. when the magnetic field inside the tubes reaches certain threshold value of about 10\textsuperscript{5} G \citep[e.g.][]{Caligari1995}.
In a simplified picture, the rising part of a flux tube forms co-called $\Omega$-loop with upper part moving to the surface, and its footpoints remaining tied to deep 
layers of the convection zone.
Crossing the visible  layer of the solar 
atmosphere (photosphere), these magnetic tubes manifest themselves in observational phenomenon called solar active regions (ARs). 
The dynamo can also act in the bulk of the CZ or even in a near-surface shear layer \citep[e.g.,][]{Krause.Raedler1980,Brandenburg2005,Stein2012,Nelson2014}. Other mechanisms may also play role 
in the formation of strong magnetic flux. For example, \citet{Brandenburg2016} showed that the 
bipolar magnetic regions can be formed due to negative effective magnetic pressure instability 
from thin magnetic flux tubes that presumably exist throughout the stratified CZ. Alternative 
mechanisms of formation of sunspots (and active regions) at/near visible solar surface by the 
photospheric dynamo had also been proposed 
\citep{Gurevich.Lebedinsky1946,Akasofu1984,Henoux.Somov1987,Henoux.Somov1991}, but currently, 
those mechanisms are unable to reproduce several major observational properties of sunspot groups 
(e.g., Hale polarity rule and latitudinal behavior of tilt angles of sunspot groups) and thus, 
are not widely accepted.

In past studies, considerable attention was paid to the observational aspects of development (emergence) of new ARs. \citet{Zwaan1985} and \citet{Brants1985} found that during the initial stage of the emergence a strong transverse magnetic field of the rising magnetic tube apex is observed. This was later confirmed by \citet{Lites1998} who also found ``a low canopy of weak horizontal magnetic field arches over the emergence zone''. As active region develops, small-scale features with magnetic fields of opposite polarities appear near its central part and move to opposite ends of the emerging flux region to form rapidly-growing footpoints of future AR \citep{Zwaan1985,Centeno2012}.
The magnetic field strength in the emerging flux elements is about several hundred gauss, and the strength increases up to kilogauss range when they become vertically oriented \citep{Lites1998}.

Recent numerical modeling indicates that when an $\Omega$-loop
reaches near surface layers, it expands 
significantly and emerges in a less coherent pattern, similar to the observed evolution of 
emerging regions.
The whole AR may be formed of several $\Omega$-loops, each one emerges individually 
\citep{Otsuji2011,Abramenko2017}. As a rule, the same polarities of these $\Omega$-loops 
gradually merge together to form the AR's footpoints. For a comprehensive observational 
description of AR's formation and evolution, see reviews by \citet{vanDrielGesztelyi2015} and 
\citet{Cheung2017}. Movie of emergence of a small active region based on Hinode 
observations can be found at \url{https://www.youtube.com/watch?v=oFF7xdcEMFg}.

The process of magnetic flux formation and rising through the CZ is inaccessible for direct observations. Helioseismology can detect rising magnetic structure inside the CZ \citep[e.g.][]{Birch2013}, however this approach still provides limited data on magnetic fields \citep{Chen2017}. This is why our knowledge on magnetic flux emergence from the bottom of the CZ almost completely relies on theoretical deductions and numerical simulations.
\citet{Schuessler1979} performed two-dimensional magnetohydrodynamic (MHD) simulations of flux 
emergence from the base of the CZ and found that the differential rotation is a suitable 
mechanism for magnetic field amplification. He also concluded that the rising flux tube 
eventually distorts and fragments into two vortex rolls. This may cancel the lifting 
force of magnetic buoyancy that was considered in detail by \citet{Longcope1996}. Studies of 
thin toroidal flux tube emergence by \citet{Choudhuri1987} and \citet{DSilva1993} showed that 
the Coriolis force may be responsible for ARs' tilt. These findings were confirmed in 3D 
numerical simulation of emerging flux loops evolution by \citet{Fan1993}. The authors also 
simulated the asymmetry of the leading and of the following polarities that is widely observed in 
real AR: the leading sunspot tends to be more coherent than the following one \citep[e.g.]
[]{McIntosh1981,Tlatov.etal2015}. Similar results were obtained by \citet{Caligari1995} who 
treated the dynamics of emerging magnetic flux tubes using the thin flux tube approximation 
proposed by \citet{Spruit1981}. \citet{Emonet1998} carried out numerical MHD simulations of a 
twisted magnetic tube rising inside a stratified medium. They found that the twist of magnetic 
tube can suppress the fragmentation of the tube into two vortex rolls described by 
\citet{Schuessler1979} and \citet{Longcope1996}.

\citet{Archontis2004} proposed the ``two-step'' emergence model of the magnetic flux tube rising from the upper layers of the solar interior into the solar corona. The model was elaborated later by \citet{Toriumi2011}. In the framework of the model, at the first step magnetic flux tube rises through the CZ due to magnetic buoyancy and slows down as it reaches just below the photosphere where the background plasma pressure falls dramatically. A horizontal layer of unmagnetized plasma trapped between the solar surface and the magnetic flux tube prevents further rising. The magnetic flux tube expands significantly in the horizontal direction, its rise halts. At the second step, horizontal sheet of magnetic flux breaks into the photosphere and further into the corona either due to magnetic buoyancy instability or due to buffeting by plasma convective motions \citep{Archontis2004,Norton2017}. \citet{Toriumi2013} predicted a divergent horizontal flow of the trapped plasma from the emerging site that was later observationally confirmed by \citet{Toriumi2014}.

\citet{Cheung2007} and \citet{Cheung2008} carried out radiative 3D MHD simulations of emerging magnetic flux from the near-surface layer of the CZ into photosphere. Magnetic flux of the simulated tubes did not exceed 1.55$\times$10\textsuperscript{20} Mx. By taking into account compressibility, radiative transfer and partial ionization, the authors simulated the flux tube undulations that were associated with interaction between the granular convection cells and the flattened magnetic sheet. 

\citet{Cheung2010} presented first radiative MHD simulations of AR emergence from the uppermost 7.5 Mm of the CZ. The simulations were carried out with the MURaM code \citep{Vogler2005, Rempel2009} that takes into account realistic equation of state. Similar study was performed by \citet{Rempel2014}, this time authors extended the model by including a retrograde flow in the flux tube and by using larger computational domain. In both works, a semitorus-shaped magnetic flux tube was advected into the computational domain through the bottom boundary. The most pronounced features of AR's emergence, such as horizontal expansion of magnetic flux tube, undulating field lines between the opposite polarities, leading and following polarities asymmetry and its coalescence from small-scale magnetic features were simulated. Interestingly, \citet{Rempel2014} showed that completely untwisted magnetic flux tube is able to rise through the uppermost 15.5 Mm of the CZ and to form sunspots.

The reader can see that most simulations consider either the uppermost layers of the CZ and solar atmosphere or the Sun's interior from the bottom of the CZ to several tens of Mm below the photosphere. This is a consequence of a density variation by 6 orders of magnitude with the CZ depth from the bottom to the top and, as a result, of significant variations of time and length scales \citep{Cheung2010}. This difficulty was overcome by \citet{Chen2017} who carried out the state-of-the-art simulations of an emergence of a magnetic flux bundle generated at the bottom of the CZ. Instead of advecting an ideal semi-torus tube through the bottom boundary of the computational domain, authors introduced magnetic and flow fields obtained near the top boundary of convective dynamo simulation by \citet{Fan2014}. This allowed \citet{Chen2017} to perform a more realistic simulation that could be compared to observable ARs, which led to the following conclusions:
\begin{enumerate}
\item{} The magnetic flux bundle rises as almost a coherent structure. Approaching the solar surface, the flux bundle fragments into small granular-size magnetic elements. The elements emerge individually and coalesce into a large flux concentration.
\item{} Asymmetry of leading and following polarities are explained by stronger vertical magnetic field of the leading spot inside the CZ that is caused by a combination of vertical motions of plasma inside rising flux tube (due to drainage from loop top) and the prograde flow inside he emerging flux bundle. 
\item{} Emergence of magnetic flux is controlled by convective upflows inside the CZ, i.e. the rising speed of a magnetic tube is determined by the mean upflow speed of plasma in convective cells.
\end{enumerate}

Additional reviews on magnetic flux tube emergence can be found elsewhere \citep[e.g.,][]{Fan2004,Fan2009,Archontis2012,Cheung2014}.

In our present study, we focus on a flux emergence rate $R(t)=d\Phi(t)/dt$, where $\Phi(t)$ is 
the total magnetic flux of the emerging flux bundle, and its relation with other magnetic 
parameters.
\citet{Otsuji2011} carried out a statistical study of 101 flux emerging regions ranging from 
small magnetic dipoles of 2.6$\times$10\textsuperscript{17} Mx to strong ARs of 6.7$\times$10
\textsuperscript{21} Mx. They found that the flux emergence rate is proportional to the peak 
total magnetic flux, $\Phi\textsubscript{max}$:
\begin{equation}
R=d\Phi(t)/dt\propto\Phi_{max}^\kappa
\label{eq1}
\end{equation}
By fitting their $d\Phi(t)/dt$ versus $\Phi\textsubscript{max}$ scatter plot, \citet{Otsuji2011} obtained that the power-law index $\kappa$ in \eqref{eq1} equals 0.57. 
The authors also derived the value of the power-law index $\kappa$ in the framework of a very simplified model. The model considered emergence of a flattened magnetic flux tube with a constant rise velocity. Since the plasma $\beta$ in the photosphere is almost 1 and the magnetic pressure $B^2/8\pi$ is almost constant, the magnetic flux density inside the tube was supposed to be constant irrespective of the spatial size or total magnetic flux of the flux tube. In such a case, the relation between the peak flux and flux emergence rate is described by equation \eqref{eq1} with the power-law index $\kappa$=0.5.


Even more emerging events, namely 224, were analysed by \citet{Khlystova2013} during the first 12 
hours of emergence. Although \citet{Khlystova2013} did not evaluated the power-law index 
$\kappa$ in her $d\Phi(t)/dt$ versus $\Phi\textsubscript{max}$ relation, a scatter plot in 
fig.~8 of her article allowed \citet{Abramenko2017} to estimate the value of $\kappa$ to be of 
about 0.4.

\citet{Fu2016} studied the influence of emerging magnetic flux on the flaring activity of the 
pre-existed ARs. They determined the flux emergence rate of 116 ARs, although did not carry out 
any analysis of these values. Interestingly, \citet{Fu2016} found that about half of the regions 
exhibited a two-step emergence behavior when the flux emergence rate was relatively low at the 
initial stages of the emergence and significantly increased as the emergence proceeded.
\citet{Norton2017} analysed the magnetic flux variations of 10 emerging ARs separately in leading 
and following sunspots. The observations also covered the decaying phase where possible. The 
authors concluded that the flux emergence rate of the leading spot in AR is on average higher 
than that of the following spot. Combining their results and others reported previously, 
\citet{Norton2017} evaluated the power-law index $\kappa$ to be 0.36. \citet{Abramenko2017} 
measured the flux emergence rate for 36 ARs. Their $d\Phi(t)/dt$ versus 
$\Phi\textsubscript{max}$ relation fitting resulted in the power-law index $\kappa=0.69\pm 0.10$ 
for the whole ARs sample.

The flux emergence rate, often reported in Mx~hr\textsuperscript{-1}, usually varies from units 
to tens of 10\textsuperscript{19} Mx~hr\textsuperscript{-1} for emerging magnetic bundles of 
different magnetic fluxes. Observed and simulated rates are in a good agreement, and we refer 
reader to an excellent detailed review of this issue made by \citet{Norton2017}. However, there 
are still some points that need to be clarified. \citet{Abramenko2017} noted that the flux 
emergence rate of ARs of almost the same total magnetic flux can differ significantly. This 
implies that the flux emergence rate also depends on some other properties of an AR rather than 
on its maximum total unsigned flux only.

The aim of this work is to find possible relations between different parameters of emerging ARs using a statistically significant sample of events. We especially focus on physical quantities that can affect the flux emergence rate of an AR. The most promising quantity in this context (except $\Phi\textsubscript{max}$) is the magnetic tube's twist. A number of theoretical studies, which simulated magnetic flux tube emergence \citep[e.g.][]{Moreno-Insertis1996, Emonet1998, Magara2001, Murray2006, Toriumi2011}, suggest that higher initial twist of a magnetic tube may result in a higher flux emergence rate. Briefly, twist of the magnetic tube prevents its fragmentation by the plasma convective flows and formation of vortex rolls mentioned above \citep{Schuessler1979, Longcope1996}. \citet{Toriumi2011} argued that insufficient twist of a magnetic tube can lead to a ``failed emergence'' when the flux tube expands too widely beneath the photosphere, looses its buoyancy and becomes unable to break through the surface. In general, \citet{Cheung2007} showed that the ability of magnetic flux tube to emerge through the surface depends on the tube's twist. Currently, there are no observational studies to support or disprove the existence of relation between the magnetic flux tube's twist and its flux emergence rate. The present study is an attempt to fill this gap.

\section{Data reduction and calculation of active-region parameters}

We employ solar magnetic field measurements by the Helioseismic and Magnetic Imager \citep[HMI,][]{Schou2012, Scherrer2012} on board the Solar Dynamics Observatory \citep[SDO, ][]{Pesnell2012}. Among other products, SDO/HMI provides full-disc 4096$\times$4096 pixel line-of-sight (LOS) magnetograms with the spatial sampling of 
0.5$\times$0.5 arcsec$^2$ pixel size
and a 720 second cadence.

To create a series of patches of AR LOS magnetic field maps we applied the following procedure. 
The boundaries of the desired AR were manually selected at the full-disc magnetogram. By a cross-
correlation technique, the AR was kept inside the bounding box at the consecutive (both back and 
forth in time) full-disc magnetograms as long as the centre of the bounding box was located at 
the central meridian distance (CMD) less than 60\degr. 

The AR's magnetic field inside the box were extracted and saved to patches along with the 
coordinates of the box boundaries. At the following steps, these coordinates were used to derive 
the information on the AR position on the solar disc.

The total unsigned flux of an AR $\Phi(t)$ was calculated as a sum of absolute magnetic flux density in pixels multiplied by pixel's area corrected for projection. We summed only those pixels whose absolute magnetic flux density values were more than 18 Mx~cm\textsuperscript{-2}, i.e. threefold exceeded the noise level of LOS 720s magnetograms \citep{Liu2012}. Assuming the magnetic field to be predominantly radial the total unsigned flux of each patch was corrected for the foreshortening by dividing by the cosine of the angle between the line-of-sight and the radial direction at the patch centre (the $\mu$-correction). The reliability of the estimation of the radial magnetic field $B_{r}$ from the LOS component $B_{los}$ by the $\mu$-correction is related to how radial the real magnetic field is. This issue was analysed by \cite{Leka2017} who compared different technique of $B_{r}$ approximation from $B_{los}$. They considered $B_{r}$ derived from the SDO/HMI vector magnetic field measurements to be a true radial field. The authors found that the $\mu$-correction provides the best estimation of the total unsigned flux in ARs from the $B_{los}$ component for $\cos(\mu)\gid0.5$ \citep[see Fig.~11 in ][]{Leka2017}. In early stages of magnetic flux emergence, the emerging site is comprised mostly of plage areas in which the magnetic field is close to radial. 
In later stages, when the horizontal field of sunspots develops, $\mu$-correction method will underestimate the 
total magnetic flux. However, based on \citet{Leka2017}, the effect of this underestimate on total unsigned flux 
of active region is relatively small.

The SDO/HMI team also provides the Spaceweather HMI Active Region Patches \citep[SHARPs,][]{Bobra2014, 
Hoeksema2014} derived from full-disc HMI data. To produce SHARPs, a feature-recognition algorithm 
\citep{Turmon2010} automatically identifies and tracks ARs at full disc magnetograms. SHARP 
products contain maps of AR's LOS and vector magnetic fields, continuum intensity, Doppler 
velocities, etc. In addition, a number of AR indices such as the total unsigned flux, 
characteristic twist parameter, mean vertical current density, and others are computed from the 
vector magnetic field in AR patches.  The SHARP vector magnetic filed data were used to calculate 
exclusively twist of magnetic flux tubes for two reasons. First, many SHARP patches enclose more 
than one AR, i.e., in such a case, we cannot attribute any AR parameter derived from this patch 
to a certain AR. Second, SHARP tracking algorithm often starts to track emerging AR after the 
actual emergence onset, i.e. we can lose information at the very beginning of the emergence.

Note that we did not use the SDO/HMI vector magnetic field data to calculate the ARs' total unsigned fluxes. The main reason is a relatively high noise in full-vector measurements. Absolute magnetic flux density below 220 Mx~cm\textsuperscript{-2} in vector magnetograms should be considered as a noise \citep{Bobra2014}. \cite{Norton2017} found that the noise in the SDO/HMI Stokes vector profiles varies with increasing centre-to-limb observing angles. This causes a ``mexican hat'' trend in the total unsigned flux curve (see fig.~5 in \citealt{Hoeksema2014} and fig.~4, panel A4 in \citealt{Bobra2014}) as an AR crosses the solar disc. \cite{Norton2017} suppressed this trend by using a high noise threshold of 575 Mx~cm\textsuperscript{-2}. However, this decision led to an underestimation of the magnetic fluxes. We cannot follow \cite{Norton2017} and increase the threshold since we intend to add to our analysis weak ARs that do not form sunspots.

There are two more minor points against using vector magnetic field measurements for $\Phi(t)$ calculations in this work. First, \cite{Leka2017} argued that the total unsigned flux derived from the vector magnetograms is overestimated due to higher noise in the transverse magnetic field. This noise is summed by its absolute value and makes some contribution to the total unsigned magnetic flux. In our case, this contribution might be significant for weak ARs. Second, as we mentioned above, SHARP tracking algorithm often starts to track an AR after its actual emergence onset. Using vector magnetic field measurements would require padding such SHARP series with additional patches in order to create a homogeneous data series covering the whole AR emergence process. To create these patches, one have to deal with full-disc vector magnetograms that consume four time more memory space compared to LOS magnetograms.

\begin{figure*}
	\includegraphics[width=\linewidth]{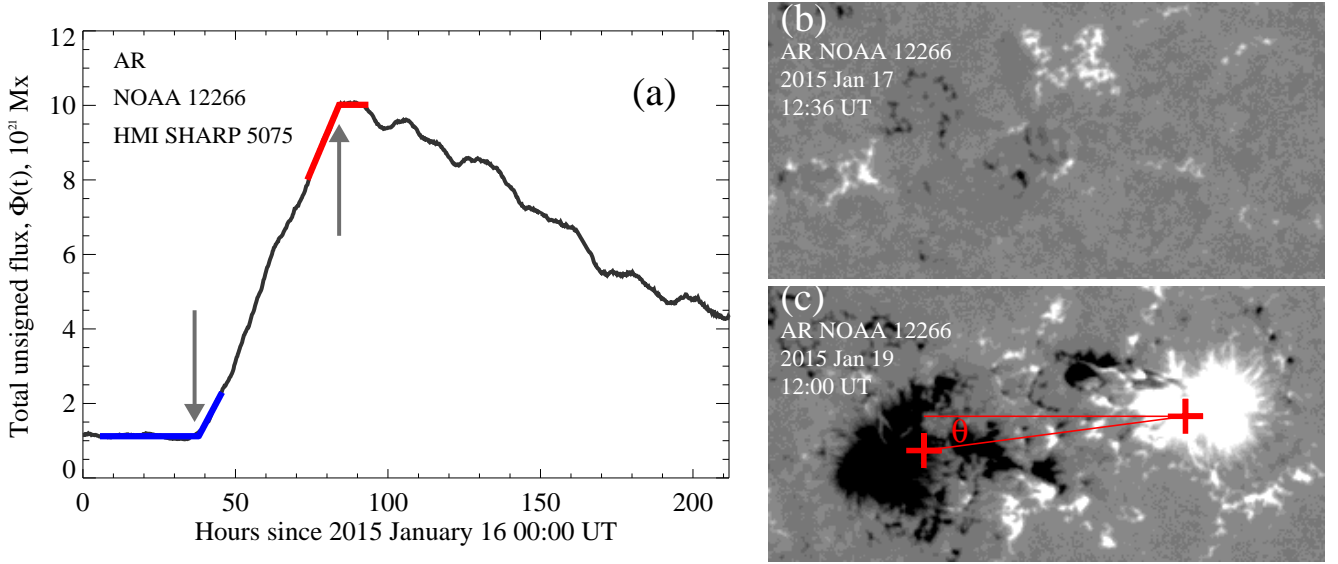}
	\caption{a - Variations of the total unsigned magnetic flux $\Phi(t)$ (black curve) of AR NOAA 12266 (HMI SHARP 5075). Blue and red curves show the best piece-wise linear fitting of the $\Phi(t)$ curve during the emergence onset and peak magnetic flux, respectively. Intersection of the segments of these fittings (pointed by gray arrows) determines $t_{emerg}$ and $t_{max}$, see text. b - LOS magnetogram of the patch of AR NOAA 12266 taken by SDO/HMI at $t_{emerg}$ on 2015 January 17 at 12:36 UT. Initial imprints of AR emergence are visible at the centre of the patch. c - LOS magnetogram of the patch of AR NOAA 12266 taken by SDO/HMI at $t_{max}$ on 2015 January 19 at 12:00 UT. Red crosses denotes the positions of flux-weighted centroids of opposite polarities. The field-of-view of the patches is 150\arcsec$\times$75\arcsec. The magnetograms are scaled from -500 Mx~cm\textsuperscript{-2} (black) to 500 Mx~cm\textsuperscript{-2}.}
\label{fig1}
\end{figure*}

Using a series of extracted by ourselves $B_{los}$ patches, for each AR we determined the following parameters:
\begin{enumerate}
\item{}
\textit{Time of the emergence onset, $t_{emerg}$.} To determine the time of the emergence onset we performed a two-segment piecewise continuous linear fitting of the $\Phi(t)$ curve in the interval where the total unsigned flux started to increase (blue curve in Fig.~\ref{fig1}a). The first (left-hand in Fig.~\ref{fig1}a) segment had to be horizontal while the second one could have arbitrary slope. We define the intersection point of the segments as a time of the emergence onset.
\item{}
\textit{The time of AR's maximum magnetic flux, $t_{max}$} was determined as a time when the emergence halts and the $\Phi(t)$ curve reaches its maximum value. Similar to the previous point, we determined this time as an intersection of two segments of the piece-wise continuous fitting of the top of the $\Phi(t)$ curve (red curve in Fig.~\ref{fig1}a). In this case, the second (right-hand) segment had to be horizontal. 

\item{}
\textit{The maximum total unsigned magnetic flux} or peak magnetic flux of an AR, \textit{$\Phi_{max}$}  was computed as a difference between the value of $\Phi(t)$ at the emergence onset and the value of $\Phi(t)$ during the $t_{max}$: $\Phi_{max}=\Phi(t_{max})-\Phi(t_{emerg})$. To decrease the uncertainty, we use the averaged values of $\Phi(t)$ at the horizontal segments of the piece-wise linear fittings.
\item{}
\textit{The averaged flux emergence rate, $R_{av}$} was computed as
the maximum total unsigned flux of an AR divided by the total interval of the flux growth
\begin{equation}
R_{av}=\Phi_{max}/(t_{max}-t_{emerg}),
\label{eq2}
\end{equation} 
\item{}
\textit{The normalized flux emergence rate, $R_{n}$}. Previous studies \citep[e.g.][]{Otsuji2011,Abramenko2017,Norton2017} 
revealed that the flux emergence rate depends on the peak flux of an AR with stronger ARs exhibiting higher absolute flux emergence rate as 
compared to weak ones. To mitigate this effect and to treat the strong and weak ARs as a single ensemble, we introduce the normalized flux 
emergence rate, $R_{n}$

\begin{equation}
R_{n}=R_{av}/\Phi_{max}^{\kappa}.
\label{eq13}
\end{equation}

$R_{n}$ is a measure of flux emergence rate that is independent of the peak magnetic flux of an AR. As it will be shown in Sec.~\ref{sec_all_vs_phimax} a value $\kappa=0.5$ can be adopted. Formally, the units of $R_{n}$ in this case are Mx\textsuperscript{1/2}~h\textsuperscript{-1}. For simplicity, we will measure $R_{n}$ in arbitrary units.
\item{}
\textit{Longitude of the emergence onset}. We adopt the heliographic Stonyhurst longitude \citep{Thompson2006} of the centre of an AR patch at the time of the emergence onset as the longitude of the emergence onset. All the heliographic coordinates were calculated with the World Coordinate System (\textsc{WCS}) routines provided in the \textsc{SolarSoft IDL}.
\item{}
\textit{Longitude and latitude of an AR at $t_{max}$, $LON_{max}$ and $LAT_{max}$}. For each AR patch we calculated the flux-weighted centroids of each polarity at $t_{max}$

\begin{equation}
(x_{\pm}, y_{\pm})=\bigg( \frac{\sum{x~B_{\pm}}}{\sum{B_{\pm}}}, \frac{\sum{y~B_{\pm}}}{\sum{B_{\pm}}} \bigg),
\label{eq3}
\end{equation}
where $B_{\pm}$ is the magnetic flux density in each pixel and $x$,$y$ are pixels' coordinates. The sign +~(-) denotes calculation only over positive (negative) polarity. To decrease the influence of quiet-Sun fields, we only used pixels with absolute values exceeding 300 Mx~cm\textsuperscript{-2}. We define the heliographic latitude and longitude of the centre of the line connecting the flux-weighted centroids as $lat_{max}$ and $lon_{max}$, respectively.
\item{}
\textit{The opposite polarity centroids separation, $d$} (or the footpoint separation between the opposite polarity centroids) is calculated as
\begin{equation}
d =  R_{\sun}\gamma,
\label{eq4}
\end{equation}
where $R_{\sun}$ is the solar radius and $\gamma$ is angular separation of two opposite polarity centroids 
(in radians, angle between two lines drawn from centre of the Sun to the centre of each centroid). 
\item{}
\textit{Tilt, $\theta$}. ARs are usually tilted with respect to the solar parallels with leading polarities tending toward the equator. This rule, first described by \citet{Hale1919}, became later known as Joy's law. Non-zero tilt angle is supposed to come from the Coriolis vortices that twist a magnetic flux bundle rising up through the CZ.
We calculate tilt of an AR as an angle between the line connecting the centroids of opposite polarities -- the axis of an AR -- and the local parallel passing through the centroid of the leading polarity (Fig.~\ref{fig1}c). Tilt is positive when measured in the counter-clockwise direction. Under these conditions, ARs obeying Joy's law will have positive (negative) tilt of units to several tens degrees in the southern (northern) hemisphere. If one adds an additional requirement that the leading polarity field should follow the Hale polarity rule \citep{Hale1925}, this would lead to a new definition of Joy's law, in which anti-Hale ARs would exhibit tilt angles of about $\pm180\degr$ \citep[cf. fig.~1a in][]{Li.Ulrich2012}.

\item{}
Finally, to compute twist, $\alpha$, of an AR, we used SHARP vector magnetic field data acquired at $t_{max}$ and remapped to Lambert cylindrical equal area projection \citep{Bobra2014,Hoeksema2014}. With $\mathbf{j}=\nabla\times\mathbf{B}$ and $\mathbf{B}$ denoting electric current density and magnetic field vector, respectively, in the framework of force-free field, the electric current could
be expressed as a product of scalar or pseudo-tensor quantity $\alpha$ and magnetic induction, {\bf B}:
\begin{equation}
\mathbf{j}=\alpha\cdot\mathbf{B}
\label{eq6}
\end{equation}
In the past, $\alpha$ was used as a measure of magnetic twist in an AR \citep{Seehafer1990,Pevtsov1994}. We calculated an averaged over the whole AR flux-weighted twist $\alpha_{av}$ \citep[e.g. ][]{Hagino2004}
\begin{equation}
\alpha_{av}=\frac{\sum j_{z}(x,y)B_{z}(x,y)}{\sum B^2_{z}(x,y)}.
\label{eq7}
\end{equation}
The summation in equation \eqref{eq7} was performed only over pixels with magnetic field strength exceeding 300 Mx~cm\textsuperscript{-2} \citep[cf. ][]{Liu2014b}.
Following \cite{Abramenko1996}, we calculated electric current density as
\begin{equation}
j\textsubscript{z}=\oint_L\mathbfit{B}\textsubscript{t}\textrm{d}\mathbfit{r},
\label{eq8}
\end{equation}
where $\mathbfit{B}\textsubscript{t}$ is the horizontal magnetic field vector (relative to 
solar surface) and $L$ is a contour that encloses an area of $5\times5$ pixels around the central point where the z-part of the electric current density $j\textsubscript{z}$ was determined.
\end{enumerate}

\section{Results and discussion}

In total, we detected more than 600 ARs emerged at the visible solar disk between 2010 May and 2017 December 31. From this initial dataset we identified 423 ARs, which satisfy the following criteria: i) both the AR's emergence onset and magnetic flux peak must be observed within 60\degr\ from the central meridian; ii) SHARP data on the AR vector magnetic field must be available at the time $t_{max}$ of the peak magnetic flux; iii) the times of the emergence onset and of peak magnetic flux can be clearly identified. In this final sample, the maximum total unsigned flux of ARs varies in the range 0.36--25.3$\times$10\textsuperscript{21}~Mx. All calculated parameters of ARs as well as its HARP numbers are listed in a table provided as a supplementary material.

\begin{figure*}
	\includegraphics[width=\linewidth]{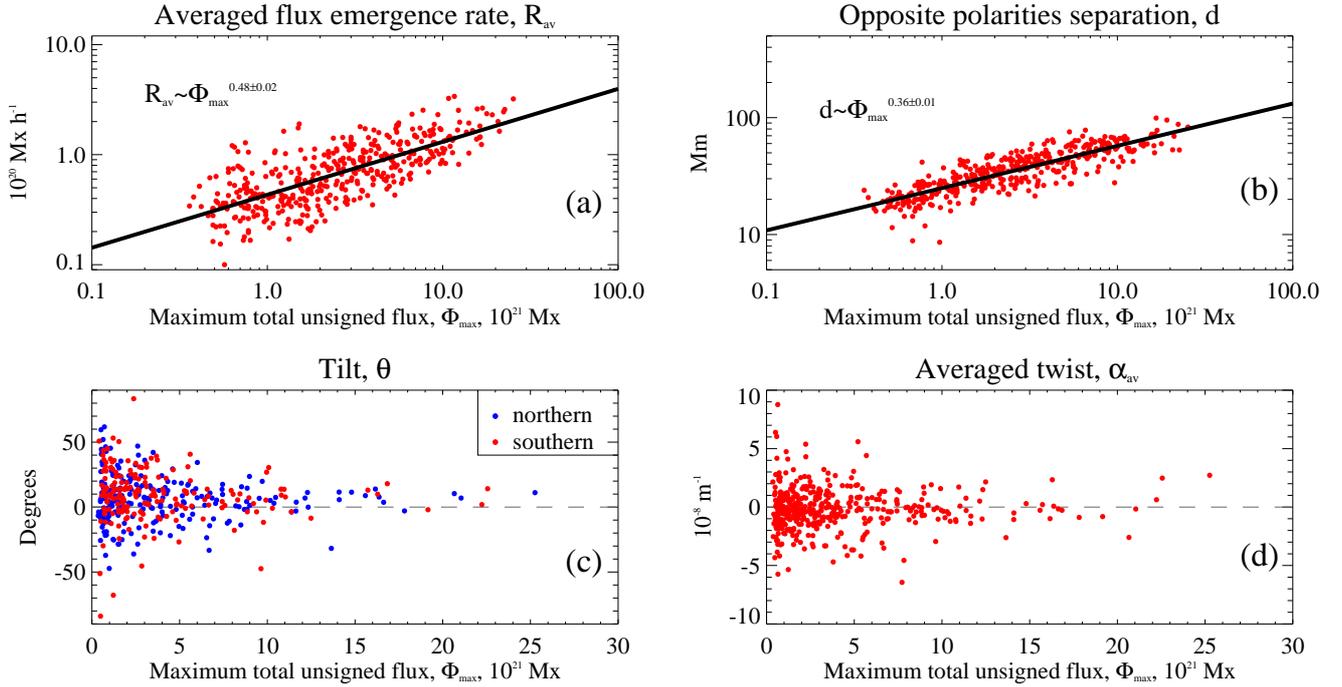}
	\caption{Relations between peak total unsigned magnetic flux $\Phi_{max}$ and averaged flux emergence rate, $R_{av}$, (a), maximum flux emergence rate, $R_{max}$, (b), opposite polarities separation, $d$ (c), ratio $R_{max}/R_{av}$ (d), tilt, $\theta$, (e), and averaged twist, $\alpha_{av}$, (f) for the entire sample of 423 ARs. Black solid lines in panels a, b, c displays the best linear fittings to the data. ARs located in the northern (southern) hemisphere are drawn in blue (red) color in panel $e$. Note that tilts for ARs located in the northern hemisphere are plotted with inverted sign in panel e. To better visualize the results, 16 outliers in panel e are excluded in the plot.}
\label{fig2}
\end{figure*}

\subsection{Dependence on maximum unsigned flux}
\label{sec_all_vs_phimax}

Fig.~\ref{fig2} shows correlation between peak total unsigned magnetic flux and other parameters. The average emergence rates as well as the polarity separation are strongly correlate with $\Phi_{max}$. Rank correlation coefficients and the probability of chance correlation are: R$_{av}$ -- 0.72 (10$^{-32}$) 
and $\rm d$ -- 0.86 (10$^{-32}$). As these two parameters are not independent of each other, strong correlation between $\Phi_{max}$ and R$_{av}$ is most likely the result of correlation between $\Phi_{max}$ and polarity separation ($\rm d$). Indeed, the R$_{av}$ residuals (not shown) after subtracting a fitted dependence between $\Phi_{max}$ and $\rm d$ (Fig.~\ref{fig2}c) do not show any correlation with $\Phi_{max}$.

Fig.~\ref{fig2}a shows a scatter plot of $R_{av}$ versus $\Phi_{max}$ in a double logarithmic scale. The best linear fitting to the distribution (black line) yields a power-law

\begin{equation}
R_{av}=0.43\Phi_{max}^{0.48\pm 0.02},
\label{eq10}
\end{equation}
where $\Phi_{max}$ is in units of 10\textsuperscript{21} Mx and $R_{av}$ is in 10\textsuperscript{20} Mx~h\textsuperscript{-1}. The power-law index is close to the theoretical value $\kappa=0.5$ obtained by \cite{Otsuji2011}. Recall that \cite{Otsuji2011} evaluated their power law index in the frame of an oversimplified emergence model assuming nearly constant magnetic flux within the tube (irrespective of the total magnetic flux or spatial size of the magnetic tube) and a constant rise velocity of the tube.

The opposite polarity centroids separation, $d$, versus $\Phi_{max}$ scatter plot is shown in Fig.~\ref{fig2}b. The best linear fitting to this scatter plot results in the following relation:
\begin{equation}
d=C\Phi_{max}^{\kappa_{1}}\propto\Phi_{max}^{0.36 \pm 0.01}.
\label{eq12}
\end{equation}
\cite{Otsuji2011} inferred that the analytical power-law index $\kappa_{1}$ in \eqref{eq12} $\kappa_{1}=2(\gamma-1)/(5\gamma-4)$, where $\gamma$ is the adiabatic index. With $\gamma=5/3$ for ideal gas and $\gamma\sim4/3$ for the near solar surface layers, $\kappa_{1}$ in \eqref{eq12} equals 0.30 and 0.25, respectively. By fitting their $d$ versus $\Phi_{max}$ scatter plot, \cite{Otsuji2011} derived the power-law index $\kappa_{1}=0.27$. Our power-law index $\kappa_{1}=0.36 \pm 0.01$ exceeds the value reported by \cite{Otsuji2011}. We suppose that this discrepancy can be explained by a different measurements criterion: \cite{Otsuji2011} defined $d=d_{max}$ as the \textit{maximum observed separation} while we consider $d$ as a distance between the opposite polarity centroids observed during the \textit{maximum observed total unsigned magnetic flux} of an AR. It was confirmed that a lag is observed between peak flux and peak separation distance in emerged magnetic bipoles, with latter trailing \citep{vanDrielGesztelyi2015,Kosovichev2008,WallaceHartshorn2012}. Of course, the mentioned discrepancy can be, at least in part, a result of specific details of AR parameter measurements in this work and in the analysis by \cite{Otsuji2011}.

Fig.~\ref{fig2}c shows the relation between tilts, $\theta$ and peak fluxes, $\Phi_{max}$. To better visualize the relation, we plotted $\theta$ with inverted sign for ARs located in the northern hemisphere (blue dots), i.e. in Fig.~\ref{fig2}c ARs obeying Joy's law have positive tilt regardless of the hemisphere where they are located. One can see that strong ARs, in general, obey Joy's law and have tilt of about ten degrees while weaker ARs exhibit scatter of tilt in a wide range of angles. The similar results were obtained by \citet{Illarionov2015} who analysed distribution of tilt angles for large and small magnetic dipoles \citep[see fig.~1 in][]{Illarionov2015}. Although tilt of an AR varies during the emergence phase \citep[e.g. ][]{vanDrielGesztelyi2015}, we can conclude that by the time of reaching its peak flux an AR has tilt expected by Joy's law.

Averaged twist of an AR, $\alpha_{av}$ versus $\Phi_{max}$ is shown in Fig.~\ref{fig2}d. There is no evident correlation between these two parameters. Nevertheless, weak ARs seem to exhibit greater absolute twist compared to strong ones. A model of helicity accumulation in magnetic flux tubes developed by \citet{Choudhuri2004} predicts that ``smaller sunspots should statistically have stronger helicity (i.e., higher values of twist''. It was observationally confirmed by \citet{Liu2014b}, however their sample consisted of only 28 emerging ARs. It seems that our results and results of \citeauthor{Liu2014b} do not contradict each other.

\subsection{Flux emergence rate versus twist}

\begin{figure}
	\includegraphics[width=\linewidth]{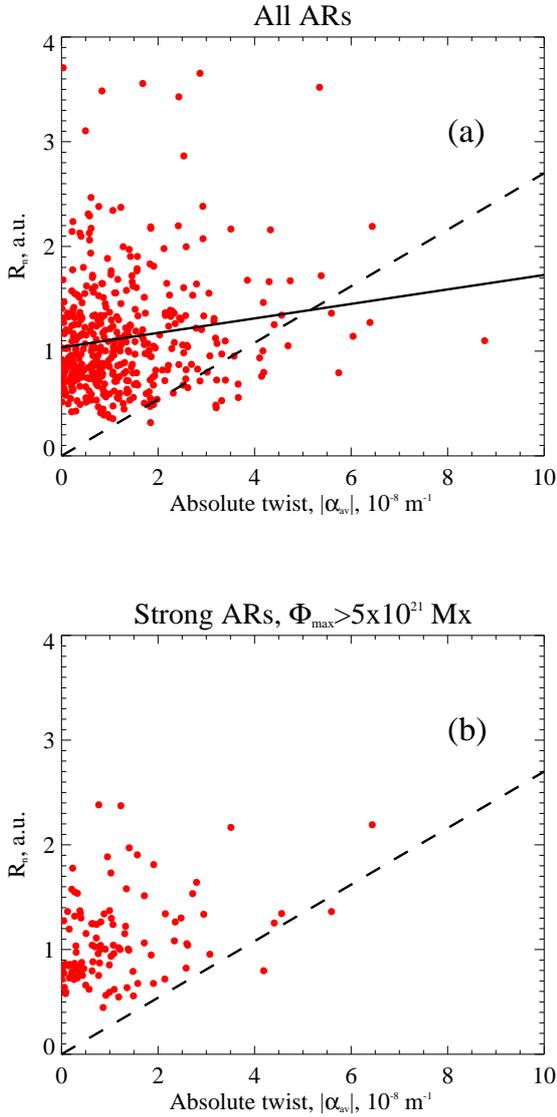}
	\caption{Relation between the normalized flux emergence rate and twist for the entire sample of ARs (a) and for strong ARs with $\Phi_{max} > 5 \times 10^{21}$~Mx (b). Black solid line in panel a displays the best linear fitting to the distribution. The dashed line in panel (b) is drawn to show that the majority of data points lie above this line. The same dashed line is overplotted in panel (a) to emphasize that the majority of points from the whole ARs sample lie above this line as well.}
\label{fig4}
\end{figure}

The $R_n$ versus $\alpha_{av}$ scatter plot for the entire sample of 423 ARs is shown in Fig. \ref{fig4}a. There 
is no obvious correlation, however, it seems that ARs emerging with low normalized rate tend to exhibit lower 
value of twist. This tendency is confirmed by the best linear fitting of the $R_n$ versus $\alpha_{av}$ relation 
(black solid line in Fig. \ref{fig4}a).

The relation between $R_n$ versus $\alpha_{av}$ shown in Fig.~\ref{fig4}a might be contaminated by influence of 
noise in the calculated twist. Indeed, the current density in weak field areas is dominated by noise due to 
uncertainties in the azimuth disambiguation procedures \citep{Hoeksema2014}. Although the thresholding keeps only 
strong field pixel during the calculation of $\alpha_{av}$ by equation \eqref{eq7}, this noise can still dominate 
at the borders of well-determined AR structures due to finite length of the integration contour in \eqref{eq8}. 
Consequently, the noise contribution to the twist value might be significant for weak small ARs.

Therefore, we further reduce our sample to 110 strong ARs with peak magnetic flux exceeding $5 \times 10^{21}$ Mx. 
In such a case, the tendency shown in Fig. \ref{fig4}a is more pronounced in the $R_n$ versus $\alpha_{av}$ 
plot for strong ARs in Fig. \ref{fig4}b. Although there is no obvious direct relation between the normalized flux 
emergence rate and twist, the majority of the data points lie above the dashed line in Fig. \ref{fig4}b 
suggesting that ARs exhibiting high twist emerge at higher flux emergence rate. This suggestion is in a general 
agreement with the theoretical predictions 
\citep[e.g.][]{Moreno-Insertis1996,Emonet1998,Magara2001,Murray2006,Toriumi2011}.

\section{Conclusions}

We used the measurements of the SDO/HMI magnetic field taken over 8 years to reveal possible relations between 
different parameters of emerging ARs. In total, our sample consisted of 423 ARs emerging and growing to its peak 
magnetic flux within 60\degr from the central meridian.

We confirmed the results of previous studies concerning the power-law relation between the flux emergence rate and peak magnetic flux of an AR. The power-law indices derived in previous works were scattered in the broad range from 0.36 in \citet{Norton2017} to 0.57 in \citet{Otsuji2011} to 0.69 in \citet{Abramenko2017}. However, the theoretical value of 0.5 obtained by \citet{Otsuji2011} was not supported observationally. In this work, using the largest so far sample of emerging ARs, we found that the power law index $\kappa$ equals 0.48$\pm$0.02 for the averaged flux emergence rate that agrees (within uncertainties) with the theoretical value of 0.5.

The opposite polarity centroids separation also scales as a power law of the peak magnetic flux, although our power law index 0.36$\pm$0.01 is higher than that obtained in the previous works. We attribute this discrepancy to different measurement routines.

Although the tilt of emerging ARs varies during the emergence phase \citep{vanDrielGesztelyi2015}, our results suggest that by the end of the magnetic flux growth an AR exhibits the tilt angle expected by Joy's law. In accordance with previous studies, the tilt angle of weak ARs is more scattered compared to strong ARs. 

The observationally obtained relation between the flux emergence rate and twist of magnetic tube shows that AR's twist sets a lower limit for the flux emergence rate. Our results suggest that more twisted magnetic tubes will emerge at higher flux emergence rate while less twisted ARs might exhibit either high or low flux emergence rate. Seemingly, there exist additional mechanisms that causes ARs of the same twist and magnetic flux to emerge with different rates.
We can suppose that those mechanisms could be attributed to the interaction between plasma convective motions and rising magnetic flux bundles. As a simple qualitative model, turbulent motions of high-density plasma inside the convection zone could fragment emerging flux bundle into several magnetic flux tubes \citep[see fig.~1 in][]{Zwaan1985}. Each tube reaches the solar surface and emerge at a different time with some delay \citep[see 'Conclusions and Discussion' section and figs.~7,8 in][]{Abramenko2017}. As a result, the duration of the whole emergence process increases leading to decrease of the averaged flux emergence rate. As another option, since the rising velocity of emerging magnetic flux is determined by the plasma upflows in convective cells \citep{Chen2017}, the emergence rate could be randomly affected by the lifting velocities of the plasma portions that carry the magnetic flux to the surface.

\section*{Acknowledgements}
The authors appreciate the referee's comments on improving this paper. SDO is a mission for NASA's Living With a Star (LWS) programme. The SDO/HMI data were provided by the Joint Science Operation Center (JSOC). The study was supported by Russian Science Foundation (Project 18-12-00131). 







\bsp	
\label{lastpage}
\end{document}